\documentclass[10pt]{article}       
\usepackage{graphicx}
\usepackage[cmex10]{amsmath}
\usepackage{amssymb,verbatim}

\def\bK{{\bf K}}

\def\mR{{\mathbb{R}}}

\def\U{{\rm U}}

\def\Tr{{\rm Tr}}
\def\Sp{{\rm Sp}}

\newenvironment{eq}
{\begin{equation}}
{\end{equation}}

\begin{document}

\title{Masking Property of Quantum Random Cipher with Phase Mask Encryption\\
-Towards Quantum Enigma Cipher-
}


\author{Masaki Sohma and     
        Osamu Hirota \\
         Quantum ICT Research Institute, Tamagawa University, \\
6-1-1 Tamagawa-gakuen, Machida, Tokyo 194-8610, Japan
}

\date{}

\maketitle
        \begin{abstract}
The security analysis of physical encryption protocol based on coherent
pulse position modulation(CPPM) originated by Yuen is one of the most
interesting topics in the study of cryptosystem with a security level 
beyond the Shannon limit.
Although the implementation of CPPM scheme has certain difficulty,
several methods have been proposed recently.
This paper deals with the CPPM encryption in terms of symplectic
transformation, which includes a phase mask encryption as a
special example, and formulates a unified security analysis for such
encryption schemes.
Specifically, we give a lower bound of Eve's symbol error probability using
reliability function theory to ensure that our proposed system exceeds the Shannon limit.
Then we assume the secret key is given to Eve  after her
heterodyne measurement. Since this assumption means that Eve has
a great advantage in the sense of the conventional cryptography, the
lower bound of her error indeed ensures the security level beyond the Shannon limit.
In addition, we show some numerical examples of the security
performance.
\end{abstract}
    \section{Introduction} 
                %
The concept of quantum random cipher was proposed by H. P. Yuen and implemented
through phase shift keying (PSK) modulation \cite{Y1} and intensity modulation (IM) \cite{H1}, which are 
called $\alpha\eta$ system or Y00 system. These systems enable us to
realize high speed direct data transmission with security protected by physical phenomena.
                Moreover he gave another implementation of quantum random cipher by using coherent pulse position modulation (CPPM) and shown that $N$-ary detection can overcome the limitation on 
the binary detection advantage of  optimal quantum receiver for PSK or IM signal states \cite{Y3}.
This brings a new scheme for both key generation and direct encryption.
%
                The CPPM system has a set of pulse position modulation (PPM) signals
\begin{equation}\label{Phi tensor}
|\Phi_\ell \rangle=|0\rangle_1\otimes\cdots\otimes|0\rangle_{\ell-1}\otimes |\sqrt{S}\rangle_\ell \otimes|0\rangle_{\ell+1}\otimes\cdots\otimes |0\rangle_N, \ell=1,....,N
\end{equation}
for $N$ messages.
                The state $|\Phi_\ell \rangle$ is encrypted to a CPPM signal
\begin{equation}\label{CPPM encryption}
|\Psi_\ell^k \rangle=U_k|\Phi_\ell\rangle=\otimes_{n=1}^N|\alpha_{\ell,n}^k\rangle_n 
\end{equation}
by a unitary operator $U_k$ randomly chosen by a running key $k$ generated from 
the pseudo random number generator (PRNG) on the secret shared key.
                The CPPM system can be realized physically by at least $N$-beam splitters \cite{Y3}, 
and  it is generally represented by symplectic transformations for multi mode Gaussian states \cite{S1}.
A defect of this system is that it does not have scalability in the implementation.
Since loss at beam splitters has a serious effect on the system, it is almost impossible 
to implement the system with enough security. So we need to find a more feasible method of
encryption.  In order to meet this requirement,
                Yuen proposed another type of encryption \cite{Y6}, where a phase mask is 
employed in order to implement the unitary operator $U_k$. The phase mask can be 
easily realized by the liquid crystal modulator (LCM) or the  acousto-optic modulator (AOM).
            %

In Shannon theory for the symmetric key cipher, 
the information theoretic security against ciphertext only attack on data has the limit
\begin{equation}\label{Shannon limit}
H(X|Y)\leq H(K),
\end{equation}
where the plaintext, the corresponding ciphertext and  the secret key are denoted by the random variables $X$,  $Y$ and  $K$ respectively.
            This is called Shannon limit for the symmetric key cipher. 
In the context of random cipher, we can exceed this limit.
Still the necessary and sufficient condition for exceeding the limit is not clear, but if the following relation holds, we can say
that the cipher exceeds the Shannon limit
\begin{equation}
H(X|Y^E,K) > H(X|Y^B,K),
\end{equation}
where the ciphertext of the legitimate receiver (Bob) and that of the eavesdropper (Eve) are denoted by 
$Y^E$ and $Y^B$ respectively.
This means that Eve cannot pin down the information bit even if she gets a secret key after measurement of the ciphertext while Bob can do it. 
We showed that the CPPM system with the encryption (\ref{CPPM encryption}) has such a property \cite{S1}.
In this paper we  clarify that our random cipher also has it.
For this purpose we evaluate Eve's symbol error probability $\bar{P}_E$ under the assumption that
Eve can know the secret key after obtaining an electrical signal by her measurement.
We give a more precise evaluation of $\bar{P}_E$ than in \cite{S1},
and obtain a lower bound of exponent $E_s(R)$ characterizing $\bar{P}_E$.
This enables us to show "strong converse to the coding theorem" \cite{Y3} for Eve's channel.
Our interest is devoted to direct encryption in particular,
but our discussions can be also applied to security analysis of 
key generation system. 
In particular the exponent $E_s(R)$ plays an important role in the latter case \cite{Y3}.
                
This paper gives a general formulation of phase mask encryption by symplectic transformations
for quantum Gaussian waveforms and presents a method for analyzing the Eve's symbol error 
probability.
            It is organized as follows: in Section \ref{Sec. 2} we overview the quantum random cipher and 
results about error probabilities of Bob. In addition we represent the PPM signals in terms of quantum Gaussian waveform as a preparation for considering the phase mask encryption in the next section. In Section \ref{Sec cannonical} we formulate the phase mask encryption basing on the general 
theory of Gaussian state developed by Holevo \cite{Holevo:82}.
In Section \ref{Sec. 4} the security of the proposed system is evaluated by
analyzing a symbol error probability of Eve.
    \section{Quantum random cipher with PPM signals}\label{Sec. 2}
        \subsection{Basic structure of quantum random cipher}
            \begin{figure}[t]
\includegraphics[width=12cm,clip]{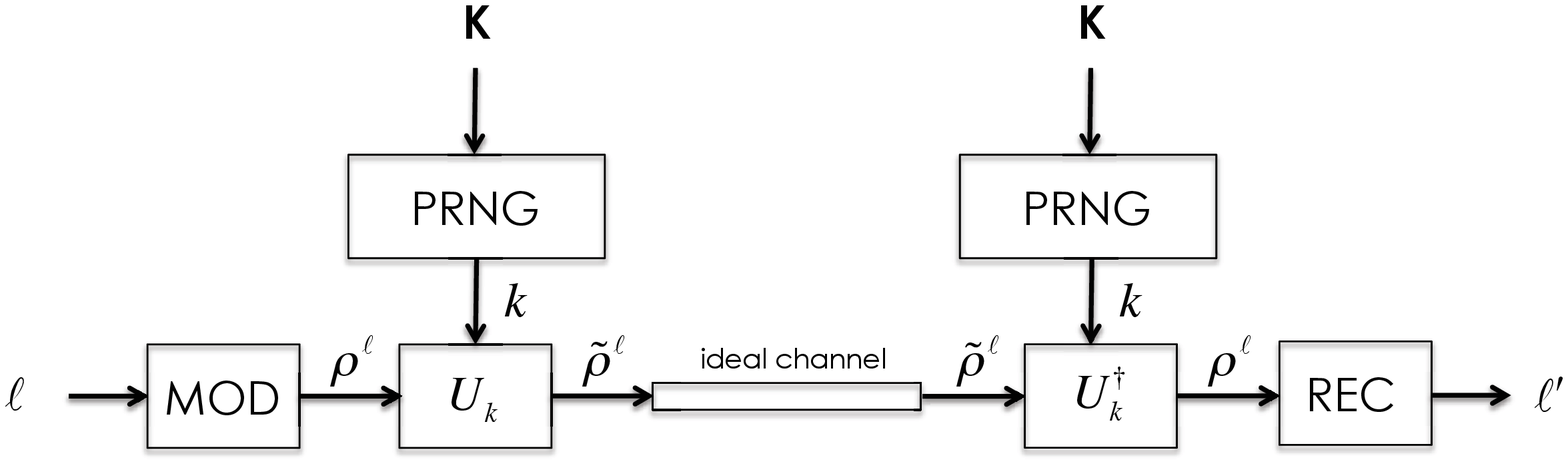}
\caption{configuration of quantum random cipher system}\label{config system}
\end{figure}
            We briefly explain a configuration of quantum random cipher (Fig. \ref{config system}). The sender(Alice) modulates her classical message $\ell$ to obtain a signal state $\rho^\ell$. 
Then the signal state is transformed into an encrypted state $\tilde{\rho}^\ell$ 
by a unitary operator $U_k$, which is randomly chosen via a running key $k$ generated
by using PRNG on a secret key {\bf  K}.
            We assume the encrypted state $\tilde{\rho}^\ell$ is sent 
through the ideal channel. Since the secret key {\bf K}, PRNG and map $k\to U_k$ are shared by Alice and Bob, Bob can apply the unitary operator $U_k^\dagger$ to the received state $\tilde{\rho}^\ell$ and obtains the signal state $\rho^\ell$.
Thus Bob can receive a classical message $\ell'$ 
with a very small error by applying the optimum detection to $\rho^\ell$.
            In contrast, Eve does not know the secret key {\bf K} and hence she must detect encrypted state $\tilde{\rho}^\ell$ directly. This makes Eve's symbol error probability worse than Bob's one.
            In the quantum random cipher, a ciphertext is protected against Eve's attack by a quantum noise.
This enables fresh key generation by communication or information theoretic security
against known plaintext attack in the symmetric key cipher.
So evaluation of Eves's symbol error probability is essential for security analysis of quantum 
random cipher.
        \subsection{Bob's symbol error probability for PPM signals}
                We restrict ourselves to the case where $\rho_\ell$ is given by the PPM signal $|\Phi_\ell\rangle\langle \Phi_\ell |$ and each mode in Eq. (\ref{Phi tensor}) is from a different
time segment. The PPM signals are quantum analog of 
orthogonal signals in classical communication theory, and error probabilities given by  
detections for them are summarized in \cite{Helstrom}.
When the optimum receiver is used, the symbol error probability is  \cite{Y7,Helstrom 2}
\begin{equation}
P_{B}^o=\frac{N-1}{N^2}\{ [1+(N-1)e^{-S}]^{1/2}-(1-e^{-S})^{1/2} \}^2.
\end{equation}  
When the photon counting receiver is used, the symbol error probability  is
\begin{equation}\label{PBc}
P_{B}^c=(1-1/N)e^{-S},
\end{equation}
where error occurs when no photons are found in any of the modes.

Let the signals be transmitted every $T$ seconds.
The signal power is $\hbar \omega_c {\cal P}=\hbar\omega_cS/T$ with an oscillator frequency $\omega_c$
and the rate is $R=\ln N/T$ [ebits/sec]. Then it is found  that for fixed any values of
${\cal P}$ and $R$ we have $P_{B}^o\to 0$ and $P_{B}^c\to 0$ as $T\to \infty$.
                This means that the capacity has an infinite value in both cases.
On the other hand,  the quasi-classical(homodyne) receiver, 
cannot achieve an infinite capacity \cite{Helstrom}.
As clarified later, Eve's optimum receiver is of such a type, and the capacity takes 
a finite value.
This implies that we can make  Eve's symbol error probability close to $1$ with keeping
$P_{B}^o$ or $P_{B}^c$ close to $0$.
The estimation of Eve's symbol error probability is shown in Section \ref{Sec. 4}.
%
        \subsection{Quantum Gaussian Waveform}
            In order to describe the unitary operator $U_k$ in Section  \ref{Sec cannonical}, we summarize the description of the electromagnetic field generated by a signal
source. For simplicity, we use the Holevo's notations given in the section IV.4 \cite{Holevo:98};  more realistic ones can be found in \cite{Yuen:78}.

                Let us consider the periodic operator-valued function
\begin{equation}
X(t)=\sum_{j=1}^\infty \sqrt{\frac{2\pi\hbar \omega_j }{T}}\left ( a_j e^{-i\omega_j t}+a_j^{\dagger}e^{i\omega_j t} \right )\quad t\in [0,T],
\end{equation}
                where $[0,T]$ is the observation interval, $a^\dagger_j,a_j$ are the creation-annihilation operators and $\omega_j=2\pi j/T$.
We assume the mode $a_j$ is described by the Gaussian states  
\begin{equation}
\rho_j(\alpha_j)=\frac{1}{\pi N_j}\int \exp\left ( -\frac{|z-\alpha_j|^2}{N_j} \right ) |z\rangle\langle z| d^2z 
\end{equation}
 with the first two moments given by  
                \begin{eqnarray}
\Tr \rho_j(\alpha_j)a_j&=&\alpha_j,\\
\Tr \rho_j(\alpha_j)a_j^{\dagger}a_j&=&N_j+|\alpha_j|^2.
\end{eqnarray}
                Then the whole process $X(t)$ is characterized by the  product Gaussian states  $\rho_\alpha=\otimes_{j=1}^\infty\rho_j(\alpha_j)$, such that  
\begin{equation}
\Tr \rho_\alpha X(t)=\alpha(t),
\end{equation}
\begin{equation}
\Tr \rho_\alpha\frac{1}{4\pi}\int_0^T X(t)^2 dt =\sum_{j=1}^\infty \hbar\omega_j(N_j+\frac{1}{2})+\frac{1}{4\pi}\int_0^T\alpha(t)^2dt.
\end{equation}
Here $\alpha(t)$ is a classical signal for quantum Gaussian channel,
\begin{equation}
\alpha(t)=\sum_{j=1}^\infty \sqrt{\frac{2\pi\hbar\omega_j}{T}}(\alpha_je^{-i\omega_j t}+\bar{\alpha}_je^{i\omega_jt}),
\end{equation}
where $\bar{\alpha}_j$ is a complex conjugate of $\alpha_j$.
                
Now let us rewrite the PPM quantum signal $|\Phi_\ell\rangle$ 
by using the representation of quantum Gaussian waveform. 
                    The classical signal corresponding to $|\Phi_\ell\rangle$ is given by
\begin{equation}\label{alpha 1}
\alpha^\ell (t)= \alpha_c(t)\chi _{I_\ell}(t)
\end{equation}
where 
\begin{equation}\label{alpha 2}
\alpha_c(t)=\sqrt{\frac{2\pi\hbar\omega_c}{T}}\sqrt{NS} (e^{-i\omega_ct}+e^{i\omega_ct}) ,
\end{equation}
\begin{eqnarray}\label{alpha 3}
\chi_{I_\ell} (t)=\left\{ \begin{array}{ll}
1, & t\in I_\ell\\
0, & t \in [0,T]\setminus I_\ell\\ 
\end{array} \right. ,
\end{eqnarray}
$\omega_c=2\pi j_c /T$ is a carrier frequency and $I_\ell = [(\ell-1)T/N,\ell T/N] $.
                    Assuming the value of $2 j_c$ is divisible by $N$, we obtain 
the energy of signal $\alpha^\ell$ as 
\begin{equation}\label{alpha energy}
\frac{1}{4\pi}\int_0^T\alpha^\ell(t)^2 dt = \hbar \omega_c S.
\end{equation}  
The Gaussian state corresponding to the classical signal  $\alpha^{\ell}(t)$ is given by
\begin{equation}\label{rho ell}
\rho^\ell = \otimes_{j=1}^\infty \rho_j(\alpha_j^\ell),\quad
\rho_j(\alpha_j^\ell)=|\alpha_j^\ell\rangle\langle \alpha_j^\ell |,
\end{equation}
                    where we obtain the values of  $\alpha_j^\ell$
from the Fourier series expansion of $\alpha^\ell(t)$:
\begin{equation}\label{classical expression}
\alpha^\ell(t)=\sum_{j=1}^\infty \sqrt{\frac{2\pi\hbar\omega_j}{T}}\left ( \alpha_j^\ell e^{-i\omega_j t}+\bar{\alpha}_j^\ell e^{i\omega_j t} \right ).
\end{equation}
                    Applying the relation
\begin{equation}\label{relation_coeff}
\alpha^\ell(t)=\alpha^1(t-\frac{\ell-1}{N}T),\quad \ell=1,....,N
\end{equation}
to Eq. (\ref{classical expression}), we have
\begin{equation}\label{relation_coeff2}
\alpha_j^\ell=\alpha_j^1e^{i\omega_j(\ell-1)T/N}=\alpha_j^1e^{i2\pi j(\ell-1)/N}.
\end{equation}
    \section{Phase mask encryption}\label{Sec cannonical}
        In this section we introduce an idea of canonical encryption basing on the general theory of
Gaussian state \cite{Holevo:82}. The canonical encryption gives a generalization of encryptions used in 
the CPPM and phase mask systems. Particularly we are interested in the phase mask 
encryption, which are formulated in the subsection \ref{subsec 3.4}.
In our phase mask encryption we apply a unitary transformation on a finite number of modes
with frequencies $\omega_j$ in the vicinity 
of a nominal carrier frequency  $\omega_c$, i.e. $j\in J=\{ j>0; |\omega_j-\omega_c| < B  \}=\{ j_1,...,j_M\}$,
although PPM signals are represented by infinite number of modes in the picture of Gaussian 
waveform. So we may also confine ourselves to a system with a finite number of degrees of 
freedom in the subsections  \ref{subsec 3.1}, \ref{subsec 3.2} and \ref{subsec 3.3}.
        Note that the canonical encryption is defined by using Stone-von Neumann theorem, 
which does not hold for an infinite number of degrees of freedom \cite{Segal 1963}.
        From Eqs. (\ref{FC1}),(\ref{F trans coeff}) and (\ref{FC2}) it is found that for an arbitrary 
small $\epsilon$ there exists $j_0>0$ such that $|\alpha_j^\ell |<\epsilon$ holds for
any $j>j_0$. So our assumption of finite bandwidth does not affect the security of 
our system. 
        \subsection{General Definition of Gaussian States} \label{subsec 3.1}
                 We give a general definition of Gaussian state.
In the following superscript $T$ denotes transpose operation for a vector or matrix. 
Let us consider the Weyl operator for a real vector $z=(z_{j_1}^q,z_{j_1}^p,....,z_{j_M}^q,z_{j_M}^p)^T$
\begin{equation}
V(z)=\exp i \sum_{j\in J}(z_j^qq_j+z_j^pp_j),
\end{equation}
where 
\begin{equation}
\begin{split}
q_j&=\sqrt{\frac{\hbar}{2\omega_j}}(a_j+a_j^{\dagger})\\
p_j&=i\sqrt{\frac{\hbar\omega_j}{2}}(a_j^{\dagger}-a_j),
\end{split}
\end{equation}
are canonical pairs satisfying the Heisenberg CCR
\begin{equation}
[q_j,p_k]=i\delta_{j,k}I,\quad [q_j,q_k]=[p_j,p_k]=0.
\end{equation}
Here $\delta_{j,k}$ takes the value of $1$ when $j=k$ and the value of $0$ otherwise.
                The Weyl operators $V(z)$ satisfy the Weyl-Segal CCR
\begin{equation}\label{WS relation}
{V}(z){V}(z^\prime)=\exp\left[ \frac{i}{2}\Delta(z,z^\prime)\right ]{V}(z+z^\prime),
\end{equation} 
                with $\Delta(z,z^\prime)=\hbar\sum_{j\in J}(z'^q_jz_j^p-z_j^qz'^p_j)$.
                The density operator $\rho$ is called {\it Gaussian} if its quantum characteristic function has the form
\begin{equation}
\Tr \rho V(z) = \exp \left[  iv^Tz-\frac{1}{2}z^TAz \right ],
\end{equation}
with mean vector $v$ and correlation matrix $A$.
%
                In particular, the Gaussian state $\rho_\ell$ given by Eq. (\ref{rho ell}) has  the mean vector
\begin{equation}
v=\Omega_M(x_{j_1},y_{j_1},.....,x_{j_M},y_{j_M})^T
\end{equation}
with $\alpha_j=x_j+iy_j$ and
\begin{equation}
\Omega_M=\oplus_{m=1}^M\left [
\begin{array}{cc}
\sqrt{2\hbar/\omega_{j_m}} &0\\
0&\sqrt{2\hbar\omega_{j_m}}
\end{array}
\right ],
\end{equation}
and the correlation matrix
\begin{equation}
A_M=\oplus_{m=1}^M\left [ 
\begin{array}{cc}
\hbar/2\omega_{j_m} &0\\
0 & \hbar\omega_{j_m}/2
\end{array}
\right ]=\frac{1}{4}\Omega_M^2.
\end{equation}
        \subsection{Symplectic Transformation} \label{subsec 3.2}
                The transformation  $L:\mR^{2M}\to \mR^{2M}$  is called {\it symplectic},
when the corresponding Weyl operators $\tilde{V}(z) =V(L^Tz)$ satisfies 
the Weyl-Segal CCR (\ref{WS relation}).
                We denote the totality of symplectic transformation by $\Sp (M,\mR )$.
                It follows from Stone-von Neumann theorem that there exists
 the unitary operator $U$ satisfying
\begin{equation}
V(L^Tz)=U^{\dagger}V(z)U
\end{equation}
for any $L\in \Sp (M,\mR)$.
                We call such derived operator $U$ the {\it unitary operator
associated with symplectic transformation}  $L$. 
%
                The characteristic function of   $\tilde{\rho}^\ell=U\rho^\ell U^{\dagger}$ is given by 
\begin{equation}
\begin{split}
\Tr\tilde{\rho}^\ell V(z)=&\Tr\rho^\ell U^\dagger V(z) U \\
=&\Tr\rho^\ell V(L^Tz)=\exp\left [ i(Lv)^Tz-\frac{1}{2}z^TLA_ML^Tz\right ]
\end{split}
\end{equation}
                In the following we confine ourselves to the Gaussian state $\rho_\ell$
given by Eq. (\ref{rho ell}), and
                our interest is devoted to the case where the state $\tilde{\rho}^\ell$ has 
the form of  $\otimes_{j\in J}|\tilde{\alpha}_j^\ell\rangle \langle \tilde{\alpha}_j^\ell |$.  
                Then the symplectic  transformation
should satisfy the condition  $LA_ML^T=A_M$, which means
\begin{equation}
\Omega_M^{-1}L\Omega_M(\Omega_M^{-1}L\Omega_M)^T=I_{2M}
\end{equation}
i.e. 
\begin{equation}\label{Omega M}
\Omega_M^{-1}L\Omega_M\in {\rm O}(2M)\cap \Sp(M,\mR)\cong \U(M)
\end{equation}  
                where  $\U(M)$ denotes the totality of  $M\times M$ unitary matrices
and ${\rm O}(2M)$  the totality of  $2M\times 2M$ orthogonal matrices.
        \subsection{Canonical Encryption}\label{subsec 3.3}
            In the {\it canonical encryption},  we encrypt the message using 
unitary operator  $U_k$ associated with  $L_k$ satisfying  
Eq. (\ref{Omega M}).
%
            In the isomorphism ${\rm O}(2M)\cap\Sp(M,\mR)\cong \U(M)$,
an element of ${\rm O}(2M)\cap\Sp(M,\mR)$,
\begin{equation}
\begin{pmatrix}
r_{11}R(\theta_{11}) & \cdots &r_{1M}R(\theta_{1M}) \\
\vdots & \ddots & \vdots \\
r_{M1}R(\theta_{M1}) & \cdots &r_{MM}R(\theta_{MM})\\
\end{pmatrix},
\end{equation}
corresponds to 
\begin{equation}\label{cann_encrypt_complex}
\begin{pmatrix}
r_{11}e^{i\theta_{11}} & \cdots &r_{1M}e^{\theta_{1M}} \\
\vdots & \ddots & \vdots \\
r_{M1}e^{i\theta_{M1}} & \cdots &r_{MM}e^{\theta_{MM}}\\
\end{pmatrix}\in \U(M).
\end{equation}
with $r_{ij}\in \mR$ and  rotation matrices $R(\theta_{ij})$.
            We denote the unitary matrix corresponding to $\Omega_M^{-1}L_k\Omega_M$
by  $U_{L_k}$. Then we can find the Gaussian state $\rho^\ell=\otimes_{j\in J}\rho_j(\alpha_j^\ell)$                                 
is encrypted into   
\begin{equation}\label{tilde rho l}
\tilde{\rho}^{\ell} =\otimes_{j\in J}\rho_j(\beta_j^{\ell}) 
\end{equation}
with
\begin{equation} \label{Encryption to beta}
(\beta_{j_1}^{\ell},.....,\beta_{j_M}^{\ell})^T=U_{L_k}(\alpha_{j_1}^\ell,.....,\alpha_{j_M}^\ell)^T.
\end{equation}     
Note that the number $M$ of encrypted modes can be taken larger than the number $N$ of pulse positions  unlike the case of CPPM with the encryption (\ref{CPPM encryption}).
        \subsection{Phase Mask Encryption} \label{subsec 3.4}
\begin{figure}[h] 
\includegraphics[width=11cm,clip]{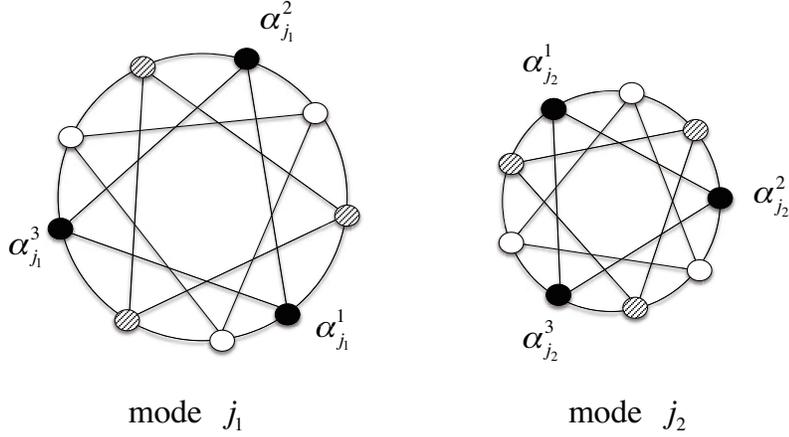}
\caption{Signal configuration of a phase mask modulation ($N=3,N'=9,M=2$)}\label{Fig_Phase_Mask}
\end{figure}

            We consider an example of the canonical encryption.
%
                If $r_{ij}=\delta_{ij}$ holds for $i,j=1,...,M$ in the matrix (\ref{cann_encrypt_complex}),  the canonical encryption is called a {\it phase mask encryption} and the matrix  (\ref{cann_encrypt_complex}) is denoted by
$U(\theta_{11},...,\theta_{MM})$.
We assume $N$ is a prime number. 
Then in the right-hand side of Eq.  (\ref{relation_coeff2}), we have
\begin{equation}
\{ e^{i2\pi j (\ell-1)/N}; \ell =1,...,N \}=\{e^{i2\pi n/N }; n=1,...,N \},
\end{equation}
if the value of $j$ is not divisible by $N$.
                So it is natural to consider the phase mask encryption given by
\begin{equation}
U_{L_k}=U(2\pi k_1/N',....,2\pi k_M/N'),
\end{equation}
where $N'$ is a multiple of $N$ 
and $k=(k_1,...,k_M)$ with $0\leq k_m< N'$ is a key generated from PRNG.
                Then each $\beta_{j_m}^\ell$ in Eq. (\ref{Encryption to beta}) takes 
values of the form 
\begin{equation}
\alpha^1_{j_m}e^{i2\pi n' /N'}, \quad n'=1,...,N'.
\end{equation}
                In particular, when $N=2, M=1$, this cryptosystem is equivalent to the $\alpha\eta$ 
cryptosystem with PSK modulation.
                In Fig. \ref{Fig_Phase_Mask}, we show an example of signal configuration 
of $\alpha_{j_m}^\ell, \beta_{j_m}^\ell$ in the case of 
$N=3, N'=9, M=2$.
The black points in the figure represent the parameters $\alpha_{j_m}^\ell$ 
for the original set of signal states $\{ \rho^\ell, \ell=1,2,3 \}$.
By the phase mask encryption with $N'/N=3$ we can use $3$ types of  sets of signal states.
                In general, the condition $p=N'/N$ gives $p$ types of sets of signal states.
%
%
%
                \begin{figure}[h]
\includegraphics[width=11cm,clip]{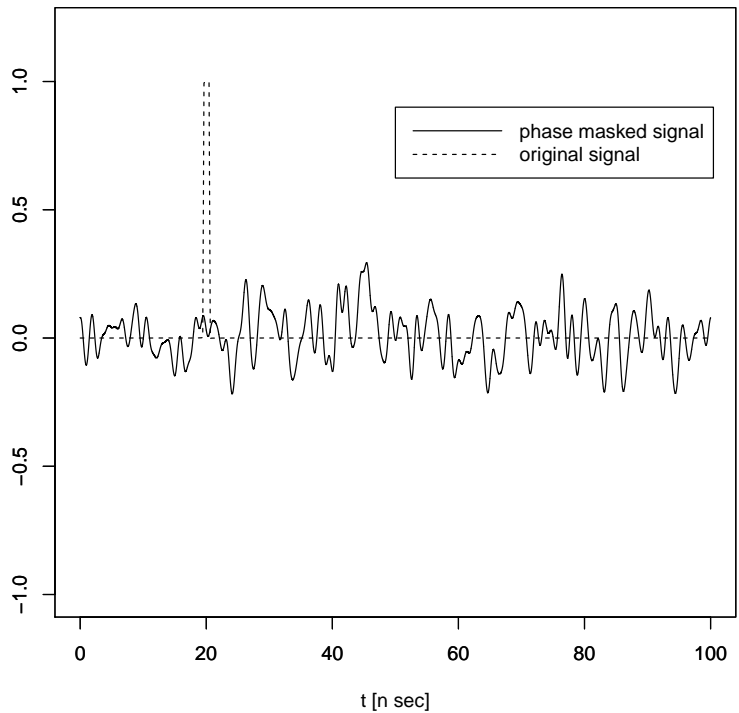}
\caption{Example of a signal transformed by the phase mask encryption}\label{signal wave fig}
\end{figure}
%

Let us obtain Fourier series for the signal $\alpha^\ell(t)$.
For simplicity we use a usual expression of Fourier series  instead of Eq. (\ref{classical expression}):
\begin{equation}\label{FC1}
\alpha^\ell(t)=\sum_{j=-\infty}^\infty c_j^\ell e^{i\omega_jt},
\end{equation}
where $c^\ell_j=\sqrt{2\pi\hbar\omega_j/T}\bar{\alpha}^\ell_j$
and $c^{\ell}_{-j}=\sqrt{2\pi\hbar\omega_j/T}{\alpha}^\ell_j$ for $j>0$.
                Then coefficients $c_j^\ell$ are computed as 
\begin{equation}\label{F trans coeff}
c_j^\ell=\sqrt{\frac{2\pi\hbar\omega_c}{T}}\sqrt{NS}(d_{j+j_c}^\ell+d_{j-j_c}^\ell),
\end{equation}
where $d_j^\ell$ are  Fourier coefficients of $\chi_{I_\ell}(t)$:
\begin{equation}\label{FC2}
d_j^\ell=\frac{1}{N}e^{-i\pi j(2\ell-1)/N}{\rm sinc} \frac{j}{N}\pi,
\end{equation} 
with ${\rm sinc} x=\sin x /x$.
                Fig. \ref{signal wave fig} shows how $\chi_{I_\ell(t)}$ is transformed by the phase mask encryption 
in a realistic setting, where
the carrier frequency is  $f_c=\omega_c/2\pi=200$[THz], the frequency resolution $1/T=10$[MHz],
the bandwidth is limited to $1$[GHz] (i.e. number of modes $M$ is $100$),
$N=97$ and $N'=9700$. In this case the transmission rate is $R=\ln N / T =45.7$ [M ebits/sec].
            %

Since  both of PPM signal and the one encrypted by a phase mask do have most of their energy at frequencies included 
in the main lobe of sinc function, effective modes for encryption are restricted 
to those frequencies. So in the case of phase mask encryption 
we cannot set the value of $M$ independently of $N$;
when we use the modes $j$ satisfying $(j-j_c)\pi/N< \pi$, $M=2N-1$.
    \section{Evaluation of Eve's  symbol error probability}\label{Sec. 4}
        \subsection{Heterodyne attack}
        Here we consider the phase mask encryption described in Sec. \ref{subsec 3.4}, avoiding complicated notations.
We can apply the same discussion to a general case.
Eve tries to estimate the classical message $\ell$ (or the secret key) from her observation of encrypted states
\begin{eq}
{\rho}(n')=\otimes_{j\in J}\rho_j(\alpha_j^1e^{i 2\pi n'_j/N'}),
\end{eq}
where $J=\{j_1,...,j_M\}$ and $n'=(n'_{j_1},...,n'_{j_M})$ with integers $1\leq n'_{j_m}\leq N'$.
        Then the error probability of her observation is given by 
\begin{eq}
P=1-\sum_{n'}\Tr{\rm \Pi}_{n'}^o \rho(n'),
\end{eq}
where $\{{\rm \Pi}_{n'}^o\}$ is a positive operator valued measure (POVM) describing Eve's optimum measurement,
and it goes to $1$ as $N'\to \infty$.
        This shows that Eve cannot estimate the message $\ell$ correctly for enough large $N'$.
However it cannot ensure that our proposed system exceeds the Shannon limit.
In order to show it, 
we have to evaluate Eve's symbol error probability $P_E$ under assumption that she can get 
the secret key {\bf K} after obtaining cipher text by the measurement $\{{\rm \Pi}_{n'}^o\}$.
        %

%
        %
In the following we evaluate Eve's symbol error probability $P_E$ under the above assumption.
In addition we may assume that she uses the following measurement on the modes $J=\{j_1,...,j_M\}$ of the states 
$\rho(n')$:
\begin{equation} \label{sub opt meas}
 X(\vec{\beta}')=\otimes_{j\in J} \frac{| \beta'_j\rangle\langle \beta'_j|}{\pi},
\end{equation} 
with $\vec{\beta}'=(\beta'_{j_1},...,\beta'_{j_M})$.
%
        Note that a measurement described by Eq. (\ref{sub opt meas}) is realized by the heterodyne detection 
when the bandwidth $2B$ is enough small \cite{Y5}.
Then Eve guesses a value of $n'$ from the result $\vec{\beta}'$of the measurement by using a function decided beforehand $g: \vec{\beta}'\to n'$. 
This gives a POVM $\{ {\rm \Pi}_{n'}\}$ describing a suboptimal measurement, where
\begin{eq}
{\rm \Pi}_{n'}=\int_{g^{-1}(n')} X(\vec{\beta}')d\vec{\beta}'.
\end{eq}
        %
Here Eve's symbol error probability $P_E$ is very little worse than that obtained by the optimum measurement$\{ {\rm \Pi}_{n'}^o \}$, 
because Eve does not know the secret key {\bf K} and there are
phase (and amplitude) uncertainties for each $\beta_j^\ell$, $j\in J$ in Eq. (\ref{tilde rho l}).
Moreover we introduce a lower bound  $\bar{P}_E$ 
of ${P}_E$
by assuming  Eve can know the secret key after her obtaining an electrical signal by the measurement (\ref{sub opt meas}).
In this case Eve can discriminate plaintext directly as explained below.
Such a condition is too advantageous to Eve and hence $\bar{P}_E$ only gives a very loose lower bound of $P_E$.
However we have to confine ourselves to evaluating the lower bound $\bar{P}_E$ 
because of a difficulty in computing $P_E$.
            %

When a message $\ell$ is transmitted, Eve obtains the signal described as a stochastic process
\begin{equation}
\beta_{E}^\ell (t)=\sum_{j\in J}\sqrt{\frac{2\pi \hbar \omega_j}{T}}\left ( {\beta'_j}^\ell e^{-i\omega_jt}+\bar{\beta'_j}^\ell e^{i\omega_j t}\right )
\end{equation}
as a result of her measurement (\ref{sub opt meas}).
            Here $\vec{\beta}'^\ell=(\beta'^\ell_{j_1},....,\beta'^\ell_{j_M})^T$ is  a random vector subject to the probability density function
\begin{equation}  \label{P ell beta}
            P_{\tilde{\rho}^\ell} (\vec{\beta}'^\ell)=\Tr \tilde{\rho}^\ell  X(\vec{\beta}'^\ell).
\end{equation}            
            In our setting, Eve knows the secret key $\bK$ after her measurement.
So using the adjoint operator $U^*_{L_k}$ of $U_{L_k}$ she can compute $\vec{\alpha}'^\ell =U^*_{L_k}\vec{\beta}'^\ell=(\alpha'^\ell_{j_1},...,\alpha'^\ell_{j_M})^T$,
which is a random vector 
            whose probability density  is 
given by $P_{\tilde{\rho}^\ell} (U_{L_k}\vec{\alpha}'^\ell)=P_{\tilde{\rho}^\ell} (\vec{\beta}'^\ell)$.
%
            From Eq. (\ref{P ell beta}), we have
\begin{equation}
\begin{split}
 P_{\tilde{\rho}^\ell} (\vec{\beta}'^\ell)=&{\rm Tr}U_k \rho^\ell U_k^\dagger X(\vec{\beta}'^\ell)\\
=&{\rm Tr}\rho^\ell U_k^\dagger X(\vec{\beta}'^\ell)U_k\\
=&\Tr \rho^\ell X(\vec{\alpha}'^\ell ) 
\end{split}
\end{equation}
            This shows that the random vector $\vec{\alpha}'^\ell$ obeys to the probability 
density function 
$\Tr \rho^\ell X(\vec{\alpha}'^\ell)$,
i.e. $\alpha'^\ell_{j_m}$ is a complex random variable with the mean $\alpha^\ell_{j_m}$ and 
the variance $1/2$.
As a result we can find that Eve obtains the signal
            \begin{equation}
\alpha_{E}^\ell(t)=\sum_{j\in J}\sqrt{\frac{2\pi\hbar\omega_j}{T}}\left ( \alpha'^\ell_j e^{-i\omega_jt}+\bar{\alpha}'^\ell_je^{i\omega_jt}\right ),
\end{equation}
where random variables $\alpha'^\ell_j$ have the mean $\alpha^\ell_j$ and the variance 
$1/2$.
        \subsection{Eve's symbol error probability}
            We evaluate Eve's symbol error probability, when she 
tries to estimate the value of $\ell$ from the signal  $\alpha^\ell_{E}(t)$.
            For the sake of brevity, we rewrite
$\alpha^\ell(t)$ and $\alpha^\ell_{E}(t)$ as
\begin{equation}
\begin{split}
\alpha^\ell(t)&=2{\rm Re}\sum_{j=1}^\infty \sqrt{\hbar\omega_j}\alpha^\ell_j\xi_j(t),\\
\alpha^\ell_{E}(t)&=2{\rm Re}\sum_{j\in J}\sqrt{\hbar\omega_j}\alpha'^\ell_j\xi_j(t),\\ \xi_j(t)&=\sqrt{\frac{2\pi}{T}}e^{-i\omega_jt},
\end{split}
\end{equation}
where we denote real part of complex number by ${\rm Re} z$.
            Let us introduce an inner product as   
\begin{equation}
\langle \psi,\varphi\rangle=\frac{1}{4\pi}\int_0^T{\psi}(t)\varphi(t)dt.
\end{equation} 
            We also use this notation for a complex function and  have 
$\langle \xi_j,\xi_k\rangle=\langle\bar{\xi}_j,\bar{\xi}_k\rangle=0$ and
$\langle \bar{\xi}_j,\xi_k\rangle=\langle {\xi}_j,\bar{\xi}_k\rangle=\delta_{j,k}$.
By definition the signals $\alpha^\ell(t)$ 
are orthogonal to each other and hence we can constitute 
the orthonormal basis  

            \begin{equation}
\psi_\ell(t)\equiv\alpha^\ell (t)/\parallel \alpha^\ell \parallel, \quad \ell=1,...,N,
\end{equation} 
where  the norm is given by
\begin{equation}
\parallel \alpha^\ell \parallel=\sqrt{\langle\alpha^\ell , \alpha^\ell \rangle}=\sqrt{\sum_{j=1}^\infty \hbar \omega_j |\alpha_j^\ell|^2}.
\end{equation}
            From Eq. (\ref{alpha energy})
we have another expression for the norm of $\alpha^\ell$ as
\begin{equation}
\parallel \alpha^\ell \parallel^2 ={\hbar \omega_c S},
\end{equation}
            which shows that $\parallel \alpha^\ell \parallel$ has a fixed value for any $\ell$.
%
            With the basis $\{\psi_\ell \}$, we represent the signal $\alpha^\ell(t)$  as the vector
$(0,...,0,\sqrt{\hbar \omega_c S},0,...,0),$
whose elements take the value  $0$  except
the  $\ell$th element.   

                Let us obtain the vector representation of $\alpha^\ell_{E}(t)$.
Its $k$th element is computed as
\begin{equation}\label{vector repr}
\langle\psi_k,\alpha^\ell_{E}\rangle
= \langle\psi_k, 2{\rm Re}\sum_{j\in J}\sqrt{\hbar\omega_j} \alpha_j^\ell  \xi_j \rangle+\langle\psi_k,2{\rm Re}\sum_{j\in J}\sqrt{\hbar\omega_j}z_j\xi_j\rangle ,
\end{equation} 
where $z_j$ is a complex random variable with the mean $0$ and the variance $1/2$. 
                Here the second term is computed as 
\begin{equation}
\frac{{\rm Re}\sum_{j\in J}\hbar \omega_j \bar{\alpha}_j^\ell z_j}{\parallel \alpha_\ell \parallel }, 
\end{equation}
which 
has the variance 
\begin{equation}
\hbar \omega_c\frac{D}{2}
\end{equation}
with
\begin{equation}
D=\frac{\sum_{j\in J}|\alpha_j^\ell|^2\omega_j^2}{\sum_{j=1}^\infty |\alpha_j^\ell|^2\omega_j\omega_c}.
\end{equation}
                Assuming that  the signals $\alpha^\ell$ have 
most of their energy at frequencies $\omega_j$ ($j\in J$) and 
 $\omega_c >> 1/(T/N)$, we have the approximations
\begin{equation}
\alpha^\ell (t)\approx 2 {\rm Re}\sum_{j\in J} \sqrt{\hbar \omega_j }\alpha_j^\ell\xi_j(t)
\end{equation}
and $d_{j+j_c}^\ell\approx 0$ in Eq. (\ref{F trans coeff}).
                Then it holds
\begin{equation}
\langle\psi_k, 2{\rm Re}\sum_{j\in J}\sqrt{\hbar\omega_j} \alpha_j^\ell \xi_j  \rangle\approx \langle\psi_k,\alpha^\ell\rangle=\delta_{k\ell}\sqrt{\hbar \omega_c S},
\end{equation}
and 

                \begin{equation}\label{DDD}
\begin{split}
D&\approx\frac{\sum_{j\in J}|\alpha_j^\ell|^2\omega_j^2}{\sum_{j\in J} |\alpha_j^\ell|^2\omega_j\omega_c}
\approx \frac{\sum_{j\in J}(\omega_cNS/\omega_j)|d_{j-j_c}^\ell|^2\omega_j^2}{\sum_{j\in J}(\omega_cNS/\omega_j) |d_{j-j_c}^\ell|^2\omega_j\omega_c}\\
&=\frac{\sum_{j\in J'}|d_{j}^\ell|^2\omega_{j+j_c}}{\sum_{j\in J'} |d_{j}^\ell|^2\omega_c}
=\frac{\sum_{j\in J'}|d_{j}^\ell|^2(j+j_c)}{\sum_{j\in J'} |d_{j}^\ell|^2j_c} \geq 1,
\end{split}
\end{equation}
                with $J'=\{ j ; | \omega_{j+j_c} -\omega_c | < B , j+j_c > 0 \}$, $\omega_j=2\pi j /T$ and $\omega_c=\omega_{j_c}=2\pi j_c /T$.
The last inequality in Eq. (\ref{DDD}) is shown by  using  $|d_j^\ell|=|d_{-j}^\ell|$ and it becomes 
the equality 
in the usual case, where
$\pm j+j_c$ 
has a positive value for any $j\in J'$.
                Thus we find the vector representation of  $\alpha_{E}^\ell$  is given by
$(w_1,w_2,..,w_{\ell-1},\sqrt{\hbar \omega_c S}+w_\ell,w_{\ell+1},...,w_N),$
where $w_k$ is a complex 
random variable with the mean $0$ and the variance $\hbar \omega_c D /2$.  In the following for simplicity we consider the equivalent situation where 
the vector representation is given by $(y_1,...,y_\ell,...,y_N)=(w_1,w_2,..,w_{\ell-1},A+w_\ell,w_{\ell+1},...,w_N)$ with $A^2=2S/D$ and $w_k$ has the mean $0$ and the variance $1$. 
%
                %

In our setting Eve uses maximum-likelihood decoding, 
where Eve picks the element $m$ for which $y_m$ is largest. 
According to \cite{Gallager}, when message $\ell$ is sent 
Eve's symbol error probability is given  by
                \begin{equation}\label{original error prob}
\bar{P}_{E}=\int _{-\infty}^{\infty} \frac{1}{\sqrt{2\pi}}\exp\left[ -\frac{(y_\ell-A)^2}{2}\right]Q_N(y_\ell)dy_\ell,
\end{equation}
where 
                $Q_N(y_\ell)$ is the probability that $y_{\ell^{\prime}}\geq y_\ell$ for some $\ell^{\prime}\neq \ell$:
\begin{equation}
\begin{split}
Q_N(y)&=1-[\Phi(y)]^{N-1},\\
\Phi(y)&=\frac{1}{\sqrt{2\pi}}\int _{-\infty}^{y} \exp(-v^2/2)dv.
\end{split}
\end{equation}
We remark that $\bar{P}_E$ does not depend on $\ell$.
                Recall the signals have duration $T$ and power $\hbar \omega_c {\cal P}=\hbar \omega_c S/T$.
Then 
\begin{equation}
A=\sqrt{2S/D}=\sqrt{2T{\cal P}/D}=\sqrt{2TC_E},
\end{equation}
with $C_E={\cal P}/D$ holds and the transmission rate is given by
\begin{equation}
R=\frac{\ln N}{T} \quad {\rm [ebits/sec]}.
\end{equation}
Let us consider a Gaussian channel, where the input power is constrained to $\hbar\omega_c{\cal P}$,
the Gaussian noise has the variance $\hbar\omega_cD/2$ and  the number of degrees of freedom is unconstrained. Then its capacity per unit time is given by
$C_E={\cal P}/D$[ebits/sec] (Corollary of Theorem 8.2.1 in \cite{Gallager}).
                In \cite{Gallager}, it is shown that $\bar{P}_{E}\to 0$ as $T\to \infty$ if $R<C_E$ 
by estimating lower and upper bounds of $\bar{P}_{E}$. We show $\bar{P}_{E}\to 1$
as $T\to \infty$ if $R>C_E$ by obtaining a lower bound of $\bar{P}_{E}$.
                    We start with the inequalities used in \cite{Gallager}:
\begin{equation}
\begin{split}
\bar{P}_{E}&\geq \int_{-\infty}^y\frac{1}{\sqrt{2\pi}}\exp\left[ -\frac{(y_\ell-A)^2}{2}\right]Q_N(y_\ell)dy_\ell\\
&\geq Q_N(y)\Phi (y-A),
\end{split}
\end{equation}
where $y$ is an arbitrary number and $Q_N(y)\Phi(y-A)$ 
is the probability that $y_\ell < y$ and 
$y_{\ell'}\geq y$ for some $\ell'$.
                    We can find lower bounds of $Q(y)$ and $\Phi(y-A)$
by using the standard inequalities on the Gaussian distribution
for $y>0$ \cite{Feller}:
\begin{equation}
\left( \frac{1}{y}-\frac{1}{y^3} \right )\frac{\exp(-y^2/2)}{\sqrt{2\pi}}<\Phi(-y)<\frac{1}{y\sqrt{2\pi}}\exp(-y^2/2).
\end{equation}
                    That is, we have
\begin{equation}
\begin{split}
Q_N(y)&=1-[1-\Phi(-y)]^{N-1}=1-\exp[(N-1)\ln (1-\Phi(-y))]\\
&\geq 1-\exp[-(N-1)\Phi(-y)]\\
&\geq 1-\exp\left[ -(N-1)\left (\frac{1}{y}-\frac{1}{y^3}\right )\frac{\exp(-y^2/2)}{\sqrt{2\pi}} \right],
\end{split}
\end{equation}
and for $y>A$
\begin{equation}
\Phi(y-A)=1-\Phi(-(y-A))\geq 1-\frac{1}{\sqrt{2\pi}(y-A)}\exp(-(y-A)^2/2).
\end{equation}
                    Putting
\begin{equation}
y=\sqrt{f \ln N}>A,
\end{equation}
we obtain the lower bound of $\bar{P}_{E}$ as
\begin{equation}\label{P_El}
\begin{split}
\bar{P}_{E}\geq &
\left(  1- \exp\left [ -\left(1-\frac{1}{N} \right) \left(1-\frac{1}{f \ln N} \right) \sqrt{\frac{N^{2-f}}{2\pi f \ln N}  }\right ]  \right ) \\
&\times \left( 1-\frac{1}{\sqrt{2\pi }(\sqrt{f \ln N}-A)} \exp \left [ -\frac{(\sqrt{f\ln N}-A)^2}{2} \right] \right )\\
=&\left(  1- \exp\left [ -\left(1-e^{-RT} \right) \left(1-\frac{1}{f RT} \right) \sqrt{\frac{\exp({(2-f)RT})}{2\pi f RT}  }\right ]  \right ) \\
&\times \left( 1-\frac{1}{\sqrt{2\pi T}(\sqrt{f R}-\sqrt{2C_E})} \exp \left [ -\frac{(\sqrt{fR}-\sqrt{2C_E})^2T}{2} \right] \right ).
\end{split}
\end{equation}
                    Here $f$ is an arbitrary positive number satisfying the inequality $\sqrt{f \ln N} > A$,
which is rewritten by using $A=\sqrt{2TC_E}$ and $N=e^{RT}$ as
\begin{equation}\label{ineq f}
f > \frac{2C_E}{R}.
\end{equation}
                    When it holds
\begin{eq} \label{CE R}
C_E<R,
\end{eq}
we can find a value of $f$ satisfying  $f<2$ and the inequality 
(\ref{ineq f}). This shows $\bar{P}_{E}\to 1$ as $T\to \infty$.
                    Putting $\bar{p}_1=1-\bar{P}_{E}$, from Eq. (\ref{P_El}) we obtain a lower bound for an exponent of $\bar{p}_1$ as 
\begin{equation}
E_s(R)=\lim_{T\to \infty}\left( -\frac{\ln \bar{p}_1}{T} \right)\geq \frac{(\sqrt{fR}-\sqrt{2C_E})^2}{2}.
\end{equation} 
Considering $f$ can take any value of $2C_E/R <f<2$, we have
\begin{equation}
E_s(R)\geq {(\sqrt{R}-\sqrt{C_E})^2}.
\end{equation}
When $T$ is large enough, $\bar{P}_E$ is approximated and lower bounded as
\begin{eq}\label{bp approx}
\bar{P}_E\approx 1-e^{-E_s(R)T}\geq 1-e^{-(\sqrt{R}-\sqrt{C_E})^2T}.
\end{eq}
                  \begin{figure}[t]
\includegraphics[width=12cm,clip]{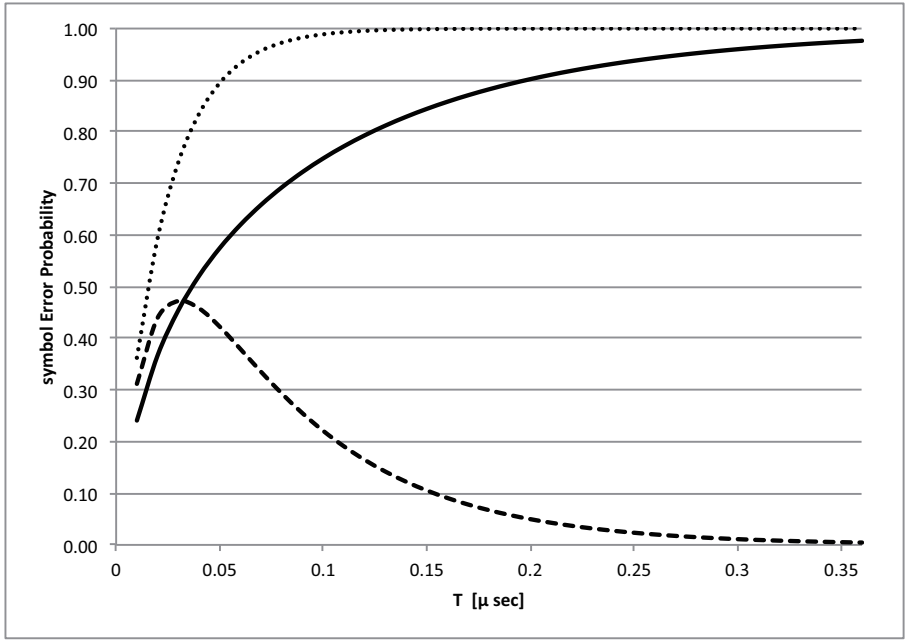}
\caption{Error probabilities of Bob and Eve with respect to signal duration $T$. Eve's symbol error probability $\bar{P}_E$ under the 
assumption that she can know the secret key after her measurement is plotted by the real line and Bob's symbol error probability $P_B^c$ the broken line. 
The dot line represents the error probability $1-1/N$ in the case where a message $\ell$ is randomly chosen from $N=e^{RT}$ messages.}\label{ErrorProbs}
\end{figure}

\begin{table}[htb]\label{table1}
  \begin{tabular}{|l||c|c|c|c|c|c|c|c|c|c|c|c|} \hline
    N & 2 & $2^2$ & $2^4$ &$2^6$  & $2^8$ & $2^{10}$ & $2^{12}$ & $2^{14}$ & $2^{18}$& $2^{22}$\\ \hline 
    T[$\mu$sec] & 0.015 & 0.031 & 0.062 & 0.092 & 0.12 & 0.15  & 0.18  & 0.21 & 0.28 & 0.34\\ \hline
  \end{tabular}
  \caption{Relation between  number of pulse positions and signal duration $T$ for the  transmission rate  $R=45$ [M ebits/sec].}
\end{table}
                Fig. \ref{ErrorProbs} gives graphs of error probabilities of Bob and Eve for 
the transmission rate $R=45$ [M ebits/sec], $C_E=15$ [M ebits/sec] and $D=1$.
In the graph the broken line represents the symbol error probability $P_B^c$ of Bob achieved by the photon counting receiver,
the solid line the lower bound $\bar{P}_{E}$ of Eve's symbol error probability $P_E$,
the dot line its upper bound $1-1/N$, which gives the error probability when a message $\ell$ is randomly chosen from $N$ messages.
The symbol error probability $\bar{P}_{E}$ is computed numerically 
from Eq. (\ref{original error prob}) without using the lower bound.
Note that $P_B^c$ has a peak  at $T=(\ln(1+R/C_E))/R=0.031$ [$\mu$ sec] and
takes a value of $4.52 \times 10^{-3}$ at $T=0.36$ [$\mu$ sec].
Here the value of $C_E={\cal P}$ is fixed and hence the pulse energy $S=C_ET$ increases as $T$ does.
So $e^{-S}$, which represents the probability that no photons are found in any of the modes, decreases 
as $T$ increases. On the other hand, when any photon is not found, we have to guess a pulse position randomly.
The error probability for such a random decision, which is given by $1-1/N=1-e^{-RT}$ in Eq. (\ref{PBc}), 
increases as $T$ does.
This is why the graph of $P_B^c=(1-1/N)e^{-S}$ has a peak
                %

The number $N$ of pulse positions is an important parameter to check feasibility of the system.
Table 1 shows a relation between $N$ and duration time $T$
for the transmission rate $R=45$ [M ebits/sec].  
By using the relations $S=TC_E$ and $\ln N=TR$, Eq. (\ref{bp approx}) and the condition  (\ref{CE R}) for it
can be respectively rewritten as
\begin{eq}
\bar{P}_E\geq 1-e^{-(\sqrt{\ln N}-\sqrt{S})^2}
\end{eq}
and 
\begin{eq}\label{eSN}
e^S<N.
\end{eq}
                Fixing the value of $E$, we find that the error probabilities $\bar{P}_{E}$, $P_B^o$ and $P_B^c$ are determined by the values of $TC_E$, $TR$.
This means that for an arbitrary number $g>0$, 
$T=(1/g)T_0$, $C_E=g C_{E,0}$ and $R=gR_0$ give the same value of the error probabilities.
On the other hand, the pulse duration  
\begin{equation}
\Delta T=\frac{T}{e^{RT}}=\frac{1}{g}\frac{T_0}{e^{R_0T_0}}
\end{equation}
takes a smaller value as $g$ (or $R$) takes a larger value.
            %

We remark on the relation between the present system and the CPPM system with the encryption (\ref{CPPM encryption}).
For the latter system we can obtain a lower bound $\bar{P}'_E$  of symbol error probability, 
assuming that Eve employs a heterodyne detection on $\alpha_{\ell,n}^k$
in Eq. (\ref{CPPM encryption})  and after her measurement she can get the secret key {\bf K} \cite{S1}.
Then it is found that the following equation holds approximately
\begin{eq}\label{P E CPPM}
\bar{P}'_E=\bar{P}_E.
\end{eq}
Here rigorously speaking it holds that $\bar{P}_E>\bar{P}'_E$ when $D>1$ in Eq. (\ref{DDD}),
but in a natural setting we have $D\approx 1$ and we may say Eq.(\ref{P E CPPM}) holds.
    \section{Conclusions}
We have formulated a security evaluation of CPPM type of quantum random
cipher in terms of quantum Gaussian waveform, and have given the
mathematical derivation process of the lower bound of Eve's symbol error
probability under the assumption that the secret key is given to Eve
after her heterodyne measurement.
This model means that Eve can try to discriminate directly plaintext
instead of ciphertext after heterodyne measurement.
Thus, this ensures a security level being beyond the Shannon limit under
stronger condition  than in the case that Eve uses the secret key after
discrimination of ciphertext based on her heterodyne measurement.
We will report a basic experiment for the latter case in the subsequent
papers.
\section*{acknowledgements}
The authors would like to thank F. Futami for his valuable discussions. 
This work was supported by JSPS KAKENHI Grant Number 24656245.

    \end{document}